\documentclass[twocolumn]{aastex631}
\usepackage{verbatim,float,xcolor,graphicx,ulem,amsmath,mathptmx}
\newcommand{\zwindow}{$0.3<z<0.4$}
\newcommand{\msun}{$M_{\odot}$}

\newcommand{\massrange}{$10^{9} \leq M/M_{\odot} \leq 10^{10}$}
\newcommand{\mbinlo}{$10^{9} \leq M/M_{\odot} \leq 10^{9.5}$}
\newcommand{\mbinhi}{$10^{9.5} \leq M/M_{\odot} \leq 10^{10}$}
\newcommand{\nsample}{575}
\newcommand{\sfr}{$M_{\odot}~$yr$^{-1}$}

\newcommand{\kmps}{km~s$^{-1}$}
\newcommand{\halpha}{H$\alpha$}
\newcommand{\lalpha}{$L_{{\rm H}\alpha}$}
\newcommand{\fha}{$\log F_{{\rm H}\alpha}$}
\newcommand{\luv}{$L_{\rm UV}$}
\newcommand{\luvo}{$L_{\rm UV}^{\rm obs}$}
\newcommand{\hbeta}{H$\beta$}
\newcommand{\deltaHa}{$\Delta\log L_{{\rm H}\alpha}$}
\newcommand{\deltaUV}{$\Delta\log L_{\rm UV}$}
\newcommand{\deltauvha}{\deltaUV~$-$~\deltaHa}

\newcommand{\avmass}{$A_V-$mass}
\newcommand{\avdev}{$A_V-\langle A_V\rangle$}

\shortauthors{Patel et al.}
\accepted{to ApJ on Feb 4, 2023}

\begin{document}

\title{Constraints on Fluctuating Star Formation Rates for Intermediate-mass Galaxies with H$\alpha$ and UV Luminosities}

\correspondingauthor{Shannon G. Patel}
\email{patel@carnegiescience.edu}

\author[0000-0003-3350-9869]{Shannon G. Patel}
\affil{The Observatories of the Carnegie Institution for Science, 813 Santa Barbara St., Pasadena, CA 91101, USA}

\author[0000-0003-4727-4327]{Daniel D. Kelson}
\affil{The Observatories of the Carnegie Institution for Science, 813 Santa Barbara St., Pasadena, CA 91101, USA}

\author[0000-0002-8860-1032]{Louis E. Abramson}
\affil{The Observatories of the Carnegie Institution for Science, 813 Santa Barbara St., Pasadena, CA 91101, USA}

\author[0000-0002-0364-1159]{Zahra Sattari}
\affil{Department of Physics and Astronomy, University of California, Riverside, 900 University Ave, Riverside, CA 92521, USA}
\affil{The Observatories of the Carnegie Institution for Science, 813 Santa Barbara St., Pasadena, CA 91101, USA}

\author[0000-0002-5337-5856]{Brian Lorenz}
\affil{Department of Astronomy, University of California, Berkeley, CA 94720, USA}
\affil{The Observatories of the Carnegie Institution for Science, 813 Santa Barbara St., Pasadena, CA 91101, USA}

\begin{abstract}

  We study the recent star formation histories (SFHs) of \nsample\ intermediate-mass galaxies (IMGs, \massrange) in COSMOS at \zwindow\ by comparing their H$\alpha$ and UV luminosities.  These two measurements trace star formation rates (SFRs) on different timescales and together reveal fluctuations in recent activity.  We compute \lalpha\ from Magellan IMACS spectroscopy while \luv\ is derived from rest-frame 2800~\AA\ photometry.  Dust corrections are applied to each band independently.  We compare the deviation of \lalpha\ and \luv\ from their respective star forming sequences (i.e., \deltaHa\ and \deltaUV) and after accounting for observational uncertainties we find a small intrinsic scatter between the two quantities ($\sigma_{\delta} \lesssim 0.03$~dex).  This crucial observational constraint precludes strong fluctuations in the recent SFHs of IMGs: simple linear SFH models indicate that a population of IMGs would be limited to only factors of $\lesssim 2$ change in SFR over $200$~Myr and $\lesssim 30\%$ on shorter timescales of $20$~Myr. No single characteristic SFH for IMGs, such as an exponentially rising/falling burst, can reproduce the individual and joint distribution of \deltaHa\ and \deltaUV.  Instead, an ensemble of SFHs is preferred.  Finally, we find that IMG SFHs predicted by recent hydrodynamic simulations, in which feedback drives rapid and strong SFR fluctuations, are inconsistent with our observations.
  
\end{abstract}


% ############################################################

\section{Introduction} \label{sec_intro}

Fluctuations in star formation histories (SFHs) record important information about galaxy formation including the supply of cold gas, metal enrichment, and morphological structures (e.g., disks and bulges) among many other properties.  Recently, some high-resolution hydrodynamic simulations have predicted intermediate-mass galaxies (IMGs) undergo bursty star formation \citep{hopkins2014} with consequences from feedback for a variety of galaxy-scale properties, including dark matter density profiles, structure, and metallicity gradients \citep{chan2015,elbadry2016}.  With order-of-magnitude fluctuations in SFR transpiring over only a few Myr in some simulations \citep[e.g.][]{sparre2017}, identifying such activity in the observations is crucial for validating prescriptions of star formation and feedback that are so crucial in modern simulations of galaxy formation.

Observationally, while measuring a current SFR is relatively straightforward, reconstructing a recent SFH (i.e., SFR as a function of time) is much more complicated.  Different methods vary with respect to the lookback time over which SFHs can be accurately recovered.  In the nearby Universe, resolved stellar populations offer perhaps the highest fidelity and highest time resolution glimpse into a galaxy's SFH \citep{weisz2014b}.  However, such an analysis is limited to galaxies in the vicinity of the Local Group.  A different method is needed to study samples in volumes large enough to characterize the full diversity of galaxy SFHs.  At higher redshifts, SED fitting to multiband photometry offers a look at the general shape of the SFH in the last few Gyr for large samples \citep[][]{kelson2014,dressler2016,iyer2017,dressler2018,leja2019}.  However, the time resolution for the most recent part of the SFH is limited by the photometric bands in the UV, which are sensitive to light from massive stars that live for $\gtrsim 100$~Myr.  This timescale is too long to capture the potentially rapid fluctuations predicted by simulations.

\begin{figure}
\epsscale{1.2}
\plotone{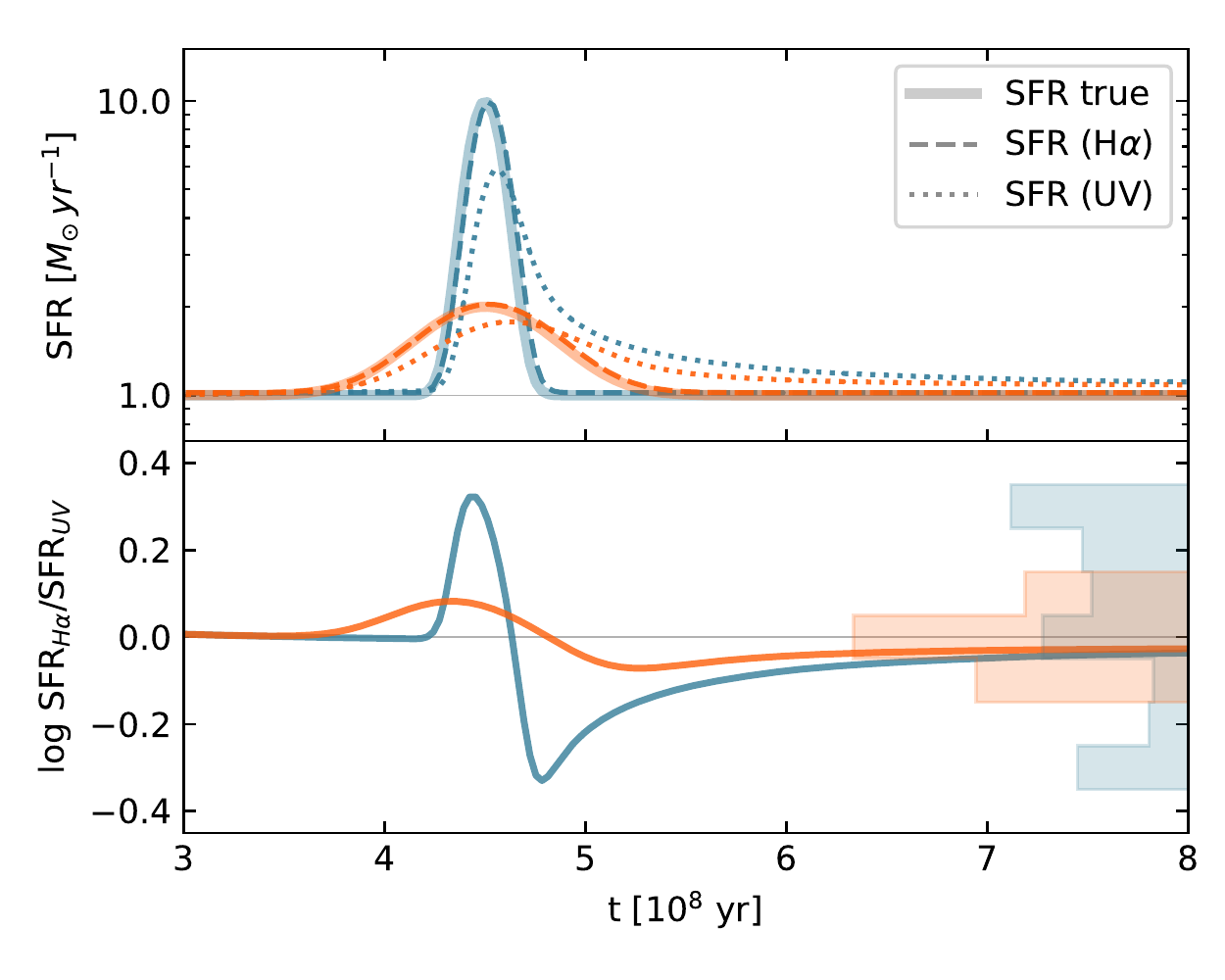}
\caption{({\em top})  Example of two fluctuating SFHs (solid curves), with one much stronger (blue) than the other (orange).  The SFRs measured from \halpha\ and UV (2800~\AA) luminosities are shown by the dashed and dotted curves, respectively.  The UV response lags \halpha\ in tracking the SFR.  ({\em bottom}) Ratio of \halpha\ to UV SFRs.  The stronger burst produces larger deviations of the ratio from unity.  Assuming the SFH characterizes a population of galaxies, sampling it uniformly in time yields a broader overall distribution of \halpha\ to UV SFR ratios (histograms on right).} \label{fig_sfr_burst}
\end{figure}

\halpha\ luminosities provide the closest measure to an instantaneous SFR, and when coupled with UV SFR tracers that register on longer timescales, can reveal the recent trajectory of an SFH \citep[see, e.g.,][]{gallagher1984, floresvelazquez2021}. As an example, Figure~\ref{fig_sfr_burst} shows two fluctuating SFHs, one with a strong burst (blue) and the other with a more gradual change in SFR (orange).  The \halpha\ and UV (2800~\AA) luminosities produced by the model SFHs are converted back into SFRs using calibrations from the literature \citep[][respectively]{bell2005,kennicutt2012}.  While \halpha\ closely tracks the instantaneous SFR, the UV indicator has longer lag times.  The bottom panel shows the ratio of the \halpha\ SFR to the UV (note that the ratio of luminosities could also be employed, as we will do in this paper).  An SFR ratio greater than unity corresponds to portions of the SFH that are elevated compared to the recent past average, while ratios less than unity are associated with diminished SFRs.  If these SFHs are representative of a population of galaxies, then sampling them uniformly in time yields a much broader SFR ratio distribution for the stronger burst, given the larger departures from unity.

\halpha\ and UV luminosities have been compared in nearby galaxies as well as at higher redshifts.  Local samples generally find larger scatter between \lalpha\ and \luv\ for lower-mass galaxies and it has been inferred that their SFHs are more bursty \citep[e.g.,][]{lee2009b,weisz2012,emami2019}.  However, these samples are small, especially in the IMG regime.  Meanwhile, high-redshift studies are mostly dependent on \halpha\ measurements from low-resolution spectroscopy \citep{broussard2019} or photometry \citep{faisst2019b}, resulting in larger systematic uncertainties.  The need for a large, high-quality sample is therefore critical, especially in the IMG regime.

In this paper, we compare \lalpha\ and \luv\ for a large sample of IMGs at low redshift.  We use their relative values to address the question of whether IMGs have smooth or strongly fluctuating SFHs.  The analysis is made possible by high-quality spectroscopic \halpha\ measurements.  The sample spans a critical stellar mass range for testing predictions from feedback models that couple strongly with SFHs.  It will also serve as a benchmark for comparison to higher-redshift samples.

The paper is structured as follows.  In Section~\ref{sec_observations} we present our observations.  In Section~\ref{sec_analysis} we compute various quantities, including reddening-corrected \lalpha\ and \luv.  In Section~\ref{sec_delta_diagram} we compare the two luminosities and determine the intrinsic scatter between them.  In Section~\ref{sec_models} we study various model SFHs in the context of our observations.  We discuss our results further in Section~\ref{sec_discussion} and we summarize our main conclusions in Section~\ref{sec_summary}.

We assume a cosmology with $H_0=70$~km~s$^{-1}$~Mpc$^{-1}$, $\Omega_M=0.3$ and $\Omega_{\Lambda}=0.7$.  Stellar masses and SFRs are based on a \citet{chabrier2003} IMF.

\section{Observations} \label{sec_observations}

\subsection{COSMOS/UltraVISTA}

Our sample of galaxies is drawn from the \citet{muzzin2013b} UltraVISTA $K_S$-selected catalog in COSMOS, which we briefly describe here.  Photometric redshifts are measured with EAZY \citep{brammer2008}.  Solar metallicity \citet[][hereafter BC03]{bc03} $\tau$-model SEDs are fit to photometry with FAST \citep{kriek2009} in up to 29 bands spanning from GALEX in the UV to {\em Spitzer} IRAC in the IR.  The SED fits produce a range of properties including stellar masses and rest-frame magnitudes at 2800~\AA\ and in $UVJ$ bands.  The SED shapes are determined by the flux within the color apertures ($D=2\farcs1$) and are scaled to a ``total'' flux by multiplying all color aperture fluxes by the ratio of the total $K_S$ flux to the color aperture $K_S$ flux.

We select IMGs from the catalog based on two simple criteria:

\begin{itemize}
\item Stellar masses between \massrange;
\item Redshifts between \zwindow.
\end{itemize}
The redshift window is motivated by a combination of the stellar mass completeness \citep[see, e.g.,][]{muzzin2013c} and the Magellan IMACS spectroscopy discussed in the next section.

\subsection{IMACS Spectroscopy} \label{sec_imacs}

We obtained Magellan IMACS \citep{dressler2011} $f/2$ multi-slit spectroscopy for a random sample of IMGs in COSMOS at \zwindow.  We used the 200~mm$^{-1}$ grism with the GG495 filter, yielding a resolution of $\sim 8$~\AA\ at $~6600$~\AA\ ($R\sim 800$).  The slit widths were $0\farcs8$.  Typical exposure times were $\sim 3$~hr, while a select few masks dedicated to $UVJ$-selected QGs were exposed for a total of $\sim 6$~hr.  The wavelength coverage enabled measurements of various strong lines, including spectrally resolved \hbeta\ and \halpha.

Kelson~et~al. (2023, in prep.) detail the reduction and extraction of the IMACS spectra as well as redshift measurements, but we provide a brief summary here.  The 2D spectra were summed uniformly along the spatial direction within $\pm 1.4\sigma$ of the object profile, resulting in a typical aperture of $\sim 1\farcs8 \pm 0\farcs6$ for the sample. We flux-calibrated the spectroscopy to match the color aperture fluxes from the photometry.  All emission-line fluxes were corrected to a total flux using the photometric aperture correction described above.  Spectroscopic redshifts were measured by cross-correlating template spectra with the observed spectra\footnote{https://code.obs.carnegiescience.edu/Algorithms/realcc}.  The redshift precision is $\sim 30$~\kmps.  A comparison to the \citet{muzzin2013b} photometric redshifts for our sample indicates the latter were precise to $\sigma_z/(1+z)\sim 0.01$.

About $\sim 11\%$ of our final sample of \nsample\ IMGs have repeat spectroscopic measurements from different masks.  These observations, which we refer to in the paper as the ``duplicate sample'', are critical for assessing accurate uncertainties in Balmer line flux measurements---and consequently in extinction-corrected \lalpha\ (Section~\ref{sec_lalpha}).

\section{Analysis} \label{sec_analysis}

\begin{figure*}
\epsscale{1.1}
\plottwo {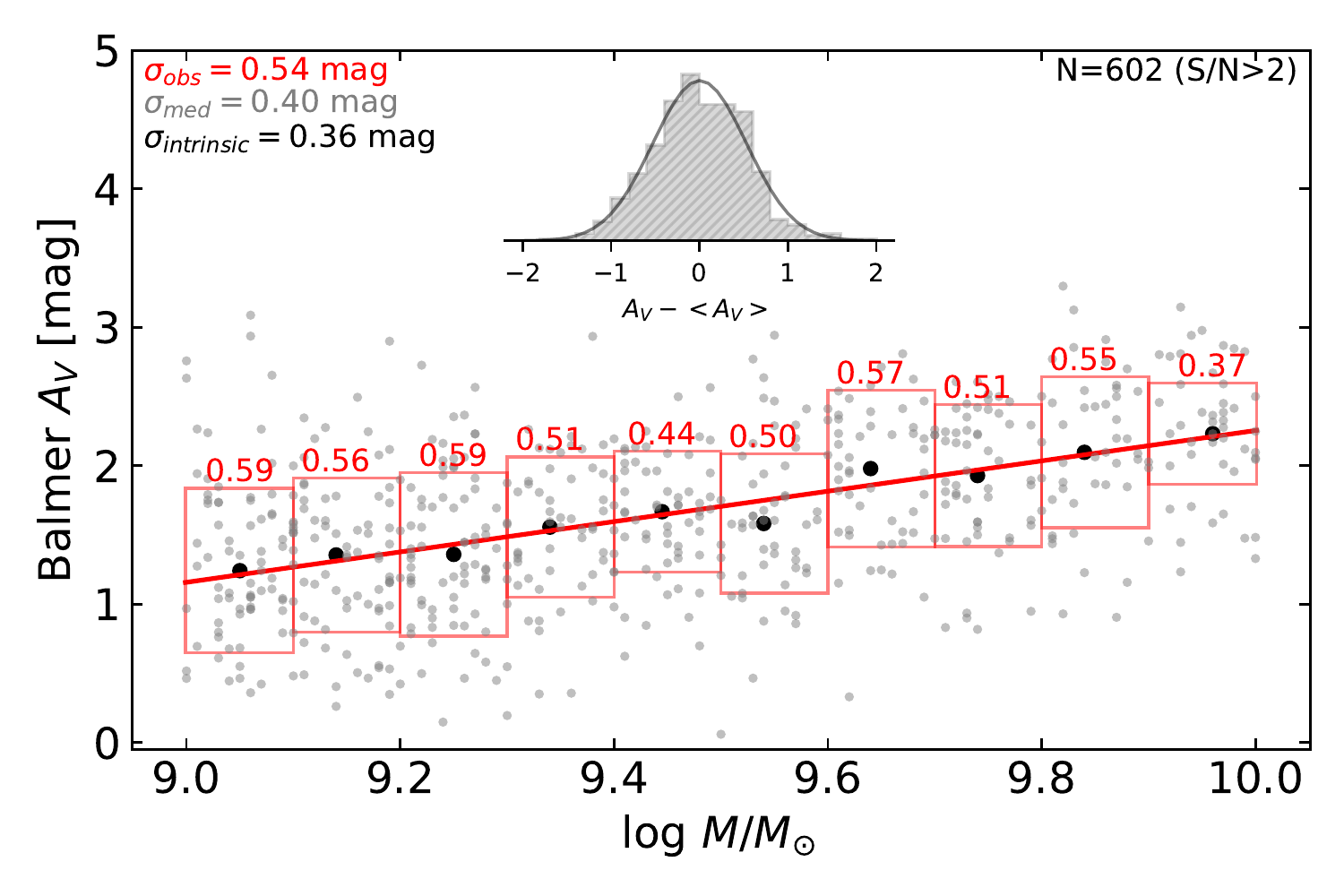}{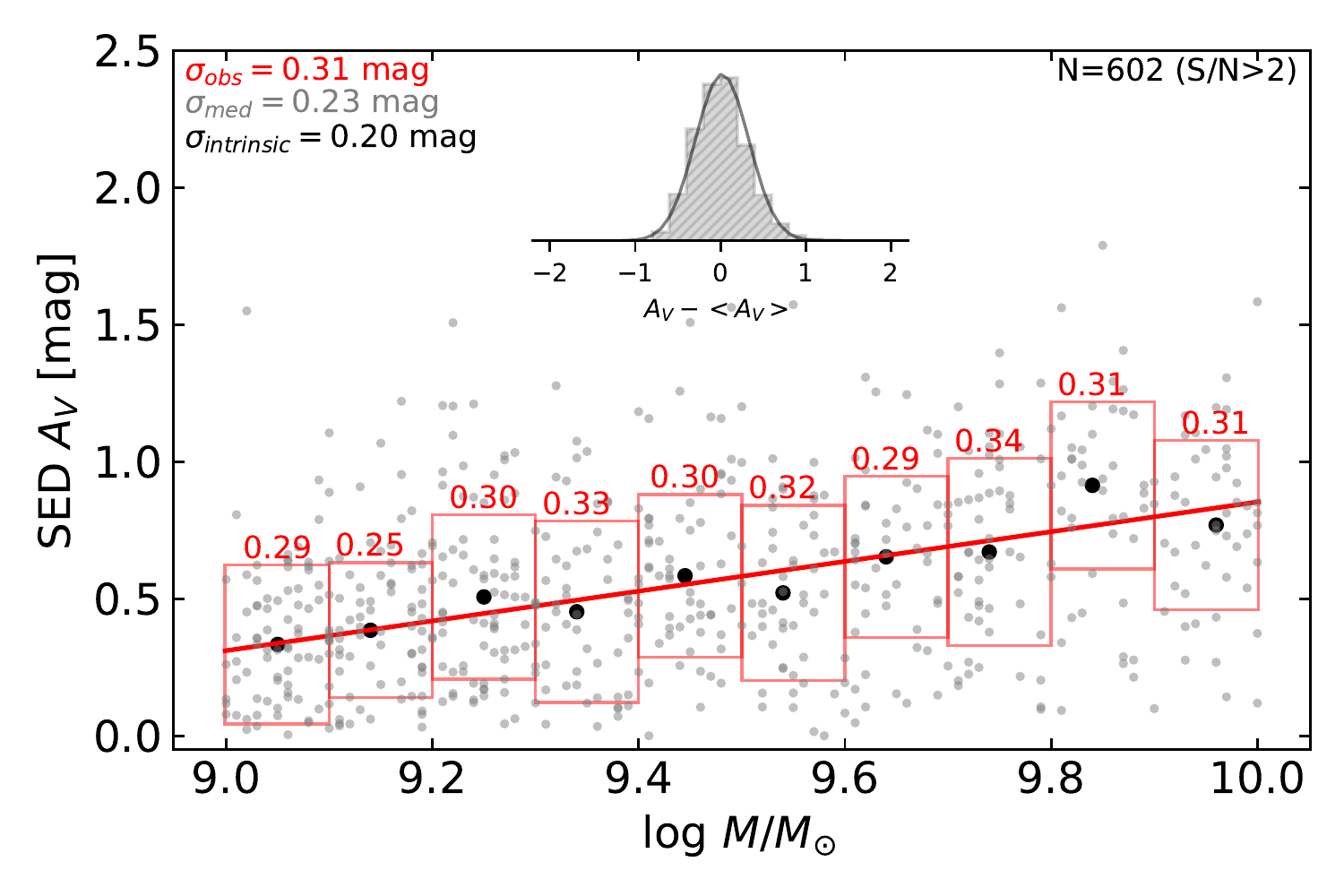}
\caption{({\em left}) Balmer $A_V$ vs. stellar mass. ({\em right}) SED $A_V$ vs. stellar mass.  For both panels, a linear fit to the median values in different stellar mass bins is shown by the red line.  The inset shows the normally distributed residuals to the fit.  Red boxes  with labels indicate the $1\sigma$ scatter about the relations, which are consistent across the mass range.  Given the observed scatters ($\sigma_{\rm obs}$) and the typical uncertainties on individual $A_V$ measurements ($\sigma_{\rm med}$), the intrinsic scatters ($\sigma_{\rm in}$) about the Balmer and SED $A_V-$~mass relations are $0.36$~mag and $0.20$~mag, respectively.  Both values are smaller than the typical individual uncertainties, indicating that a more precise measure of $A_V$ can be derived for both the SED and Balmer $A_V$ by adopting these relations.} \label{fig_avmass}

\end{figure*}

\subsection{UV Luminosity (\luv)}

The UV luminosity is one of two measurements that we use to gauge the recent SFHs of IMGs.  It traces star formation activity on timescales of $\sim 10$~Myr to $\sim 100$~Myr \citep{floresvelazquez2021}.  We use the rest-frame 2800~\AA\ luminosity ($l_{\nu,2800}$) computed by \citet{muzzin2013b}\footnote{The column labeled \texttt{L\_UV} in the \citet{muzzin2013b} catalog is equivalent to $\nu\l_{\nu,2800}$, as defined in \citet{bell2005} and adopted here.} to estimate \luv, as defined in \citet{bell2005}:
\begin{eqnarray}
L_{\rm UV}^{\rm obs}=1.5\nu\, l_{\nu,2800}
\end{eqnarray}
\luvo\ is the observed (i.e., attenuated) UV luminosity that is directly observed. The rest-frame 2800~\AA\ flux is computed with EAZY \citep[see, e.g.,][]{brammer2011}, which uses the flux directly from a linear combination of best-fitting templates to the entire SED.  For comparison, using the observed $u$-band (which samples rest-frame 2800~\AA\ at \zwindow), without interpolating between adjacent bands or applying a $K$-correction, produces similar rest-frame 2800~\AA\ fluxes with only $\sigma=0.03$~dex scatter.  Given this small scatter, we adopt the \citet{muzzin2013b} EAZY-derived $l_{\nu,2800}$ measurements with the $u$-band uncertainties for the error estimate for \luvo.  The $3\sigma$ $u$-band flux limit corresponds to a UV luminosity at $z \sim 0.4$ of $10^{41.9}$~erg~s$^{-1}$.  The entire sample is above this limit.

We correct the observed \luvo\ measurements for dust attenuation using a \citet{calzetti2000}  reddening law and estimates of $A_V$ from fitting SEDs to the starlight. To avoid degeneracies between derived values of $A_V$ and assumptions about any underlying functional forms of SFHs, we used a maximally diverse set of nonparametric SFHs realized from $H=1$ stochastic processes \citep{kelson2014,kelson2020}, in which metallicities were allowed to vary:

\begin{eqnarray}
  \log L_{\rm UV}&=&\log L_{\rm UV}^{\rm obs} + 0.4\, A_{2800} \\
  A_{2800}&=&(7.26/4.05)\,A_V
\end{eqnarray}
\luv\ is the extinction-corrected UV luminosity.  $A_{2800}$ is the extinction at 2800~\AA.  We add an additional uncertainty of $0.20$~mag to our error estimate for $A_V$.  This is consistent with the scatter between different $A_V$ derivations from SED fits that use different SFHs (e.g., fBm vs. $\tau$-models).  The median uncertainty for SED derived $A_V$ is $\sigma_{A_V}\sim 0.23$~mag.  In a subsequent section, we look at correlations between these estimates of $A_V$ and stellar mass, and we use comparisons of the scatter in this correlation with uncertainties in $A_V$ to guide how we proceed with our analysis of the UV and emission-line star formation indicators.

\subsection{\halpha\ Luminosity (\lalpha)} \label{sec_lalpha}

The \halpha\ luminosity is the other measurement that we use to study the recent SFH of IMGs, tracing star formation activity on much shorter timescales of $\sim 5$~Myr \citep{floresvelazquez2021}.  For a given spectrum, all emission lines are fit with a Gaussian with a common width and redshift.  This method enables us to flag lines with poor measurements due to contamination from nearby night sky emission.  The underlying spectrum of a galaxy’s stellar populations---including the Balmer absorption---was adopted from the nonparametric model SFHs that best fit the broadband SEDs, convolved to the resolution of the IMACS spectroscopy.  We find that the $1\sigma$ uncertainty in Balmer absorption results in variations in the final dust corrected \halpha\ emission of $<10\%$ for most of the sample - a level far below that of the more dominant dust correction.

We determine the amount of nebular extinction by comparing the observed Balmer line ratios to those expected for gas at densities and temperatures consistent with \ion{H}{2} regions \citep[\halpha:\hbeta:H$\gamma$:H$\delta$ = 1:0.35:0.16:0.09,][]{osterbrock2006}.  We assume Case B recombination as well as a \citet{calzetti2000} reddening law.  While many works measure the Balmer decrement using only \halpha\ and \hbeta, we also utilize H$\gamma$ and H$\delta$ when available and carry out a Bayesian likelihood analysis.  The unattenuated \halpha\ flux (\fha) and dust extinction ($A_V$) are the two parameters of the model.  The range of \fha\ probed varies for each galaxy and depends on the attenuated \halpha\ flux, while the extinction was probed over $0 < A_V < 10$.  Given the potential for low-S/N Balmer lines resulting in uncharacteristically high $A_V$ estimates, we implement a prior for $A_V$.  The prior was determined by modeling the summed likelihoods of the objects with more tightly constrained confidence intervals from an initial ``flat-prior'' iteration of the likelihood analysis.  To determine the measured value of \fha\ and $A_V$, we marginalize the posterior distribution over the other parameter and take the 50th percentile value to represent the parameter of interest.  The $1\sigma$ uncertainty is taken to be the difference between the 84th and 16th percentile values divided by two.

The error in the extinction correction depends sensitively on the error in the observed Balmer line fluxes, especially that of \halpha\ and \hbeta.  We examine our duplicate observations to assess uncertainties.  Among high-$S/N$ duplicates, where photon noise is negligible, observed \halpha\ and \hbeta\ fluxes both show scatters of $\sim 0.045$~dex.  This extra error needs to be accounted for when computing Balmer decrements.  Some of the extra error, however, is correlated between the Balmer lines (e.g., due to variations from slit losses and/or reddening at different slit positions) and is therefore subtracted prior to propagation through the Bayesian analysis.  We determine the amount of correlated error to subtract by comparing the change in \hbeta\ fluxes to the change in \halpha\ fluxes among high-$S/N$ duplicates.  The scatter about the correlation is $\sim 0.03$~dex, indicating that the remaining error of $\sim 0.03$~dex is uncorrelated.  This extra uncorrelated error is added in quadrature with the existing formal errors on the Balmer lines when comparing their relative values.  The source of the extra error is not entirely known.  Some portion likely originates from fitting emission lines, especially redshifted \halpha, in the vicinity of  bright skylines.

After executing the Bayesian likelihood analysis, we find the typical uncertainty for the Balmer derived $A_V$ is $\sim 0.40$~mag.  Despite inflating the errors on the Balmer line flux measurements based on well-informed assumptions, this uncertainty is still less than the scatter in $A_V$ among duplicates of $0.53$~mag. (This value is somewhat uncertain, given the smaller duplicate sample size, and it also varies depending on how it is measured, e.g., median absolute deviation, standard deviation, etc.).  We explore the implications of this difference in the next section.  The attenuated \halpha\ flux limit ($2\sigma$) is $4.0\times 10^{-17}$~erg~s$^{-1}$~cm$^{-2}$.  The corresponding limiting luminosity at $z=0.4$ is $\sim 2.3\times 10^{40}$~erg~s$^{-1}$.

\subsection{More Precise Extinction Corrections from $A_V$ vs. Stellar Mass Relations} \label{sec_avmass}

The individual SED and Balmer $A_V$ uncertainties are fairly large ($0.23$~mag and $0.40$~mag, respectively).  The uncertainties in $A_V$ propagate into errors in SFR. In turn, these limit the precision to which we can constrain SFR fluctuations on timescales of $\sim 10$-100~Myr.

The correlation between $A_V$ and stellar mass \citep[e.g.,][]{garn2010b} could be utilized to improve on our extinction uncertainties if the intrinsic scatter in the relation is less than that of the individual measurement errors.  In Figure~\ref{fig_avmass}, we use a linear fit to the median values of $A_V$ in different mass bins to represent the \avmass\ relation. For consistency, we utilize the same sample for analyzing both SED and Balmer $A_V$\footnote{The sample used in Figure~\ref{fig_avmass} requires S/N~$>2$ \halpha\ detections and is larger than our final sample of \nsample\ IMGs because joint \luv\ and \lalpha\ luminosity limits have not yet been imposed, as they require this step to be completed.  The difference in sample size is small and therefore does not impact our final results.}.  Note that the two relations follow different trends, as nebular extinction is higher than that of stellar continua \citep[e.g.,][]{calzetti2000}.  The residuals to the fit, which are normally distributed, bolster the view that much of the variation in the individually measured $A_V$ is tied to a trend about stellar mass.

The observed scatter about the Balmer \avmass\ relation is  $\sigma_{\rm obs}=0.54$~mag (all scatters reported here represent $1.48\times$~MAD) and fairly consistent across the IMG mass range.  However, much of the scatter is due to measurement uncertainties ($\sigma_{\rm med}=0.40$~mag).  Subtracting this component in quadrature leaves an intrinsic scatter of $\sigma_{\rm in}=0.36$~mag.  Assigning Balmer $A_V$ values based on the trend therefore results in a significant improvement in their uncertainties, as the intrinsic scatter becomes the formal error.  For example, the 75th percentile error of the individually derived Balmer $A_V$ values is coincidentally also $\sim 0.54$~mag.  A revision down to $0.36$~mag by adopting the intrinsic scatter above represents a sizable improvement for a large portion of our sample.  We note that our duplicate sample suggests somewhat larger measurement errors of $0.53$~mag---similar to the observational scatter in the relation and suggestive of an intrinsic scatter of $\sigma_{\rm in} \approx 0$~mag.  While acknowledging a level of uncertainty in the intrinsic scatter of the Balmer \avmass\ relation, we adopt our measured value above $\sigma_{\rm in}=0.36$~mag for the remainder of the paper, but we also show below that our conclusions remain unchanged when focusing on values from the duplicate sample.

For the SED \avmass\ relation, the observed scatter is $\sigma_{\rm obs}=0.31$~mag.  Subtracting the typical SED $A_V$ uncertainties of $\sigma_{\rm med}=0.23$~mag, the intrinsic scatter in the relation is $\sigma_{\rm in}=0.20$~mag.  Assigning SED $A_V$ values based on the trend therefore also results in an improvement in their uncertainties.

With improved uncertainties for Balmer and SED $A_V$ values through their respective \avmass\ relations, the median errors for \luv\ and \lalpha\ become $0.15$~dex and $0.13$~dex, respectively.  We utilize the \avmass\ trends as our fiducial method for extinction corrections and will discuss this choice in further detail in Section~\ref{sec_pref}.  We will also compare the results to luminosities corrected with the individual SED and Balmer $A_V$ values to check for consistency in Section \ref{sec_delta_diagram}.

\begin{figure}
\epsscale{1.2}
\plotone{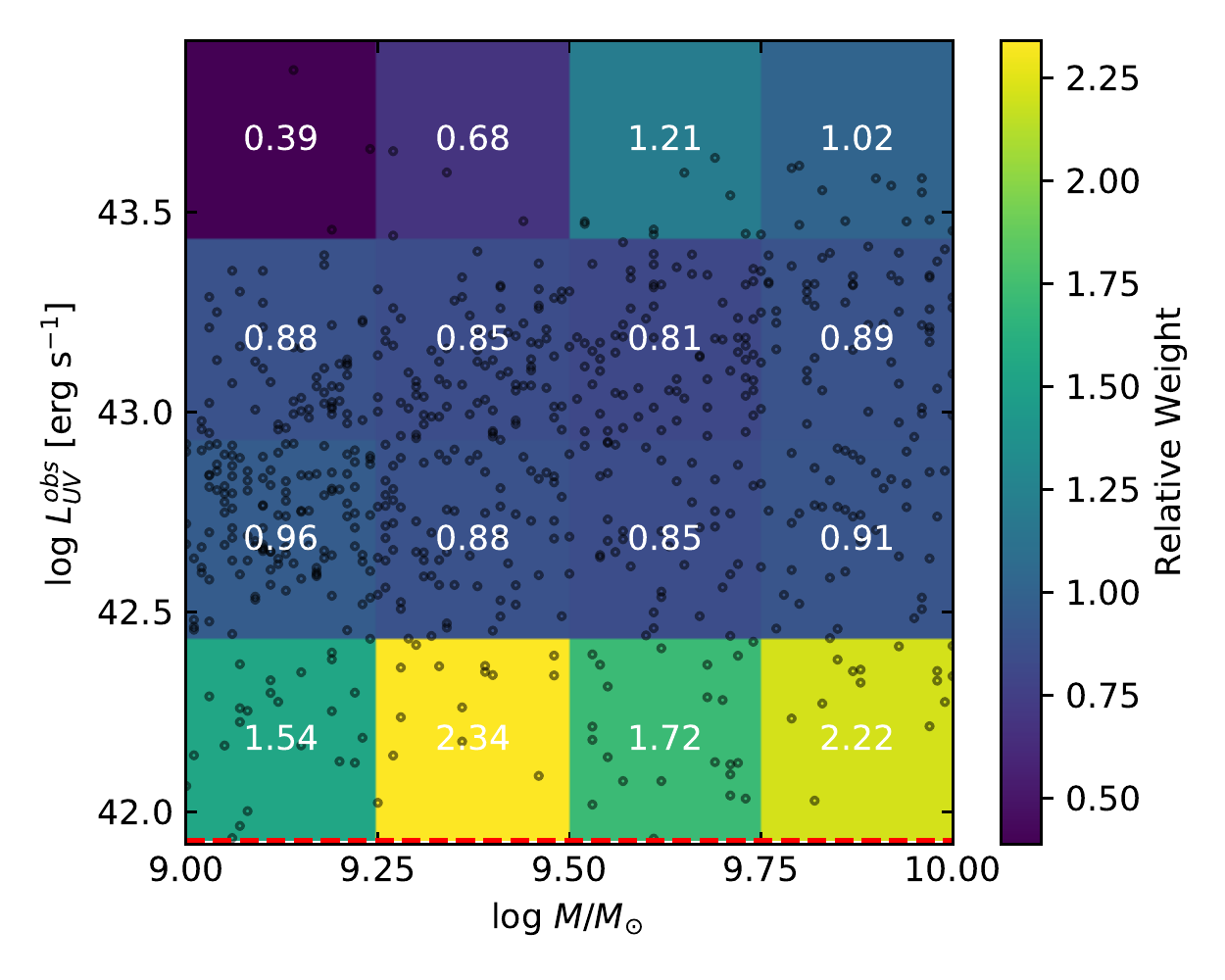}
\caption{Weights assigned to galaxies based on the completeness of our IMACS spectroscopy in COSMOS as a function of \luvo\ (i.e., the observed extincted UV luminosity) and stellar mass for IMGs at \zwindow.  Values are indicated for each bin.  Open circles indicate our sample with measured Balmer decrements.  Weights are higher at lower \luvo\ values, indicating the difficulty in obtaining spectroscopic redshift measurements for lower-mass quiescent galaxies.  The dashed red line indicates the $3\sigma$ detection limit for \luvo.} \label{fig_completeness}
\end{figure}

\subsection{Completeness}

\begin{figure*}
\epsscale{1.2}
\plotone{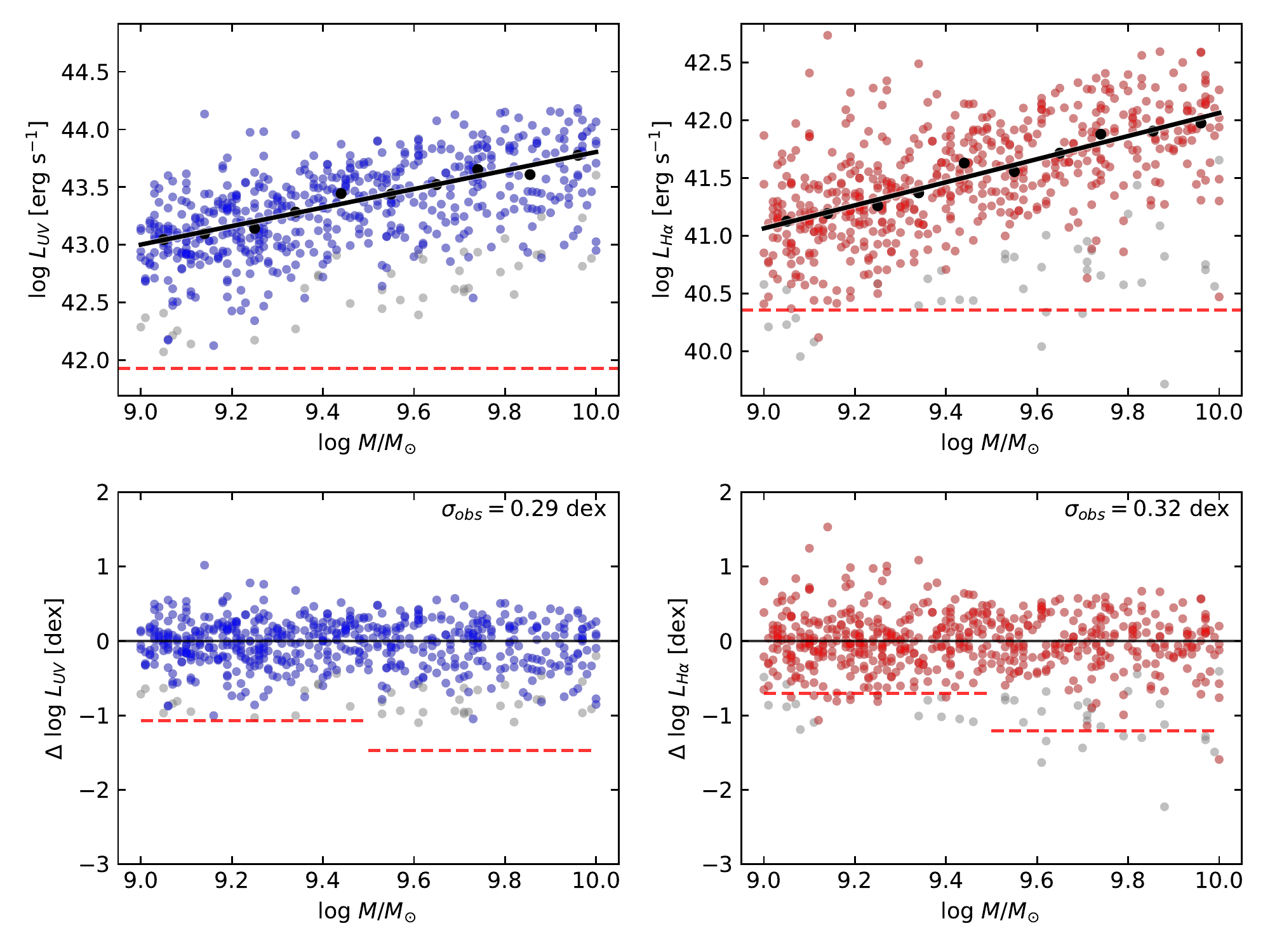}
\caption{Variations in recent SFHs are inferred from \deltaUV\ and \deltaHa, which are derived based on deviations of \luv\ and \lalpha\ from their respective SFS.  The top panels show \luv\ ({\em left}) and \lalpha\ ({\em right}) vs. stellar mass with $UVJ$-selected SFGs marked in blue and red, respectively (quiescent galaxies in gray).  Median luminosities in bins of stellar mass for SFGs are indicated by the black circles.  The black lines are linear fits to the median binned data (black circles) and represent the SFS for each SFR indicator.  Residuals (i.e., \deltaUV\ and \deltaHa) are shown in the bottom panels.  Detection limits are indicated by the red dashed lines, with the luminosity limit applied to two different mass bins in the bottom panels.} \label{fig_delta_sfs}
 \end{figure*}

 To account for luminosity-dependent completeness in our sample, we assign weights to each galaxy when necessary.  While \luv\ is measured for all galaxies in our sample, \lalpha\ is limited by spectroscopic completeness.  We parameterize the completeness of our IMACS spectroscopy as a 2D function of stellar mass and \luvo\ by comparing the subsample to the photometrically selected parent sample of IMGs at \zwindow\ in the \citet{muzzin2013b} catalog (Figure~\ref{fig_completeness}).  Accurate spectroscopic redshift measurements determine the relative completeness levels in this 2D space: identifying the redshift enables a measurement for \lalpha.  Redshift measurements are impacted by ($i$) galaxy faintness and ($ii$) the presence of spectral features that aid redshift determination.  For this reason, stellar mass and \luvo\ provide a useful basis for characterizing the completeness of our survey.  The weights are computed by normalizing the inverse completeness of a given bin by the overall inverse completeness of the survey.  Due to the relatively constant nature of the completeness at fixed \luvo, lower-mass quiescent galaxies generally have higher weights because they are more difficult to measure redshifts for, given their noisy spectral features in our data.

\section{Comparison of \luv\ and \lalpha\ for IMG\lowercase{s}} \label{sec_delta_diagram}

\subsection{Computation of \deltaUV\ and \deltaHa}

\begin{figure*}
\epsscale{1.2}
\plotone{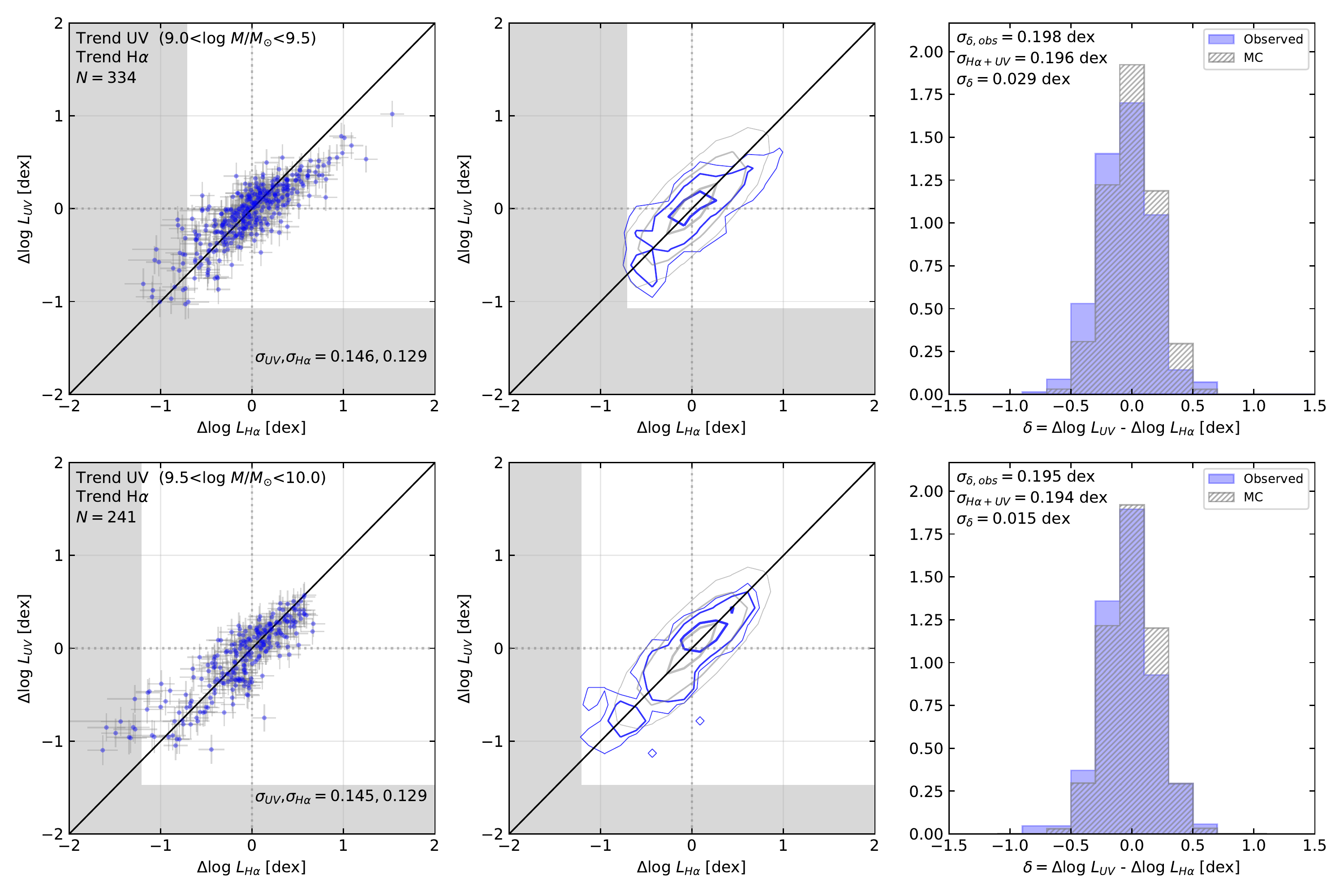}
\caption{({\em left}) Comparison of \deltaUV\ vs. \deltaHa\ for low ({\em top}) and high ({\em bottom}) mass IMGs.  \luv\ and \lalpha\ are corrected for extinction with $A_V$ derived from the SED and Balmer \avmass\ trends, respectively.  Median formal errors ($\sigma_{\rm UV}$, $\sigma_{{\rm H}\alpha}$) labeled. ({\em middle}) Contours (blue) containing 25, 75, and 95 percent of the data above the detection limits (gray shaded regions). ({\em right}) Distribution of \deltauvha\ (i.e., $\delta$) with labels for the observed scatter ($\sigma_{\delta, {\rm obs}}$), median error ($\sigma_{{\rm H}\alpha+{\rm UV}}$), and intrinsic scatter ($\sigma_{\delta}$) in $\delta$.  Also shown in the right two columns are the contours and histograms (gray) for Monte Carlo simulations that bootstrap uncertainties from the data.  Most of the observed scatter is accounted for by measurement errors.  The small intrinsic scatter between \deltaUV\ and \deltaHa\ therefore constrains the strength of SFH variations.} \label{fig_HaUV_compare}
\end{figure*}

In practice, reframing each luminosity in terms of its deviation from the star-forming sequence (SFS) (i.e., \deltaHa\ and \deltaUV) has advantages over the absolute quantities, \lalpha\ and \luv.  The slope of the SFS is known to vary for different SFR indicators, likely reflecting systematic differences in dust corrections \citep{salim2007}.  By shifting reference frames from absolute to relative luminosities, we are able to bypass such complexities in the absolute luminosity derivations.  Systematic trends of stellar metallicity---and their impact on the number of ionizing photons---with stellar mass also get taken out.  Finally, a systematic error in $A_V$, even one that is correlated with stellar mass, would have no impact on the relative luminosity quantities.

The top panels of Figure~\ref{fig_delta_sfs} show \luv\ and \lalpha, both from the subsample with \halpha\ measurements, as a function of stellar mass.  We use the BPT diagram \citep{bpt1981} to find and remove galaxies coincident with AGN spectral emission \citep[e.g.,][]{kewley2001b}.  This cut removes only $\sim 3\%$ of the sample and does not impact our conclusions.  Median luminosities are computed in bins of stellar mass for $UVJ$-selected SFGs  \citep[see, e.g.,][]{patel2012} and a linear fit (in log-log space) to these values is used to define the SFS.  We note that using all galaxies would not significantly impact the slope, as QGs are few in proportion and median binning is resistant to such outliers. The residuals from the SFS, \deltaUV\ and \deltaHa\ are shown in the bottom panels.  We use the \luv\ and \lalpha\ luminosity limits to specify detection limits for \deltaUV\ and \deltaHa\ for two mass bins spanning \mbinlo\ and \mbinhi\ (red dashed lines).  Among $UVJ$-selected SFGs, the observational scatter about the SFS is similar between the two indicators ($\sigma_{\rm obs}\sim 0.3$~dex).  Note that, although only SFGs are used to define the SFS, we can compute \deltaUV\ and \deltaHa\ for all galaxies in our spectroscopic sample.

\subsection{The \deltaUV\ vs. \deltaHa\ Diagram}

We examine \deltaUV\ vs. \deltaHa\ with two different methods for extinction corrections.  In Section~\ref{sec_delta_trend}, we present results with our fiducial method that utilizes the SED and Balmer \avmass\ trends to correct for attenuation. In Section~\ref{sec_delta_ind}, we use the individual SED and Balmer $A_V$ values.  In Section~\ref{sec_pref}, we summarize why the former is our preferred method.

\subsubsection{\luv\ and \lalpha\ Corrected for Extinction with $A_V$--Mass Trends} \label{sec_delta_trend}

Figure~\ref{fig_HaUV_compare} shows \deltaUV\ and \deltaHa, derived in Figure~\ref{fig_delta_sfs}, for two mass bins.  \luv\ and \lalpha\ have been corrected for attenuation using the \avmass\ trends from Figure~\ref{fig_avmass}.  The measurement errors in this figure ($\sigma_{\rm UV}$, $\sigma_{{\rm H}\alpha}$) are therefore mostly driven by the intrinsic scatter in those relations.  One of the advantages of examining the \halpha\ and UV distributions in this 2D space is the ability to easily apply detection limits (gray shaded region), which become more complicated when studying the ratio of the two luminosities.  The one-to-one relation (solid line) indicates where \deltaHa~$=$~\deltaUV.  Ignoring measurement uncertainties for a moment, IMGs below the relation have \lalpha\ more elevated with respect to the SFS than \luv, while above the relation \luv\ is more elevated.  As we will see in Section~\ref{sec_models}, these respective regions correspond to SFHs with elevated or diminished SFRs compared to the recent past average.  However, measurement uncertainties prevent such a definitive characterization for most galaxies in our sample.

We focus on the overall width of the joint distribution of \deltaHa\ and \deltaUV\ to gauge fluctuations in IMG SFHs.  The middle column shows blue contours containing 25\%, 75\%, and 95\% of the data (thick to thin lines) above the detection limits.  Weights from our completeness map have been applied here.  The distributions are peaked at the origin (i.e., on the SFS) and spread along the one-to-one relation. The right column shows histograms of the difference $\delta=$~\deltauvha\ for the data above the detection limits.  Crucially, the intrinsic scatter ($\sigma_{\delta}$) of this distribution has implications for the allowable strength of recent SFH fluctuations and on specific timescales.  After subtracting in quadrature the median uncertainty for $\delta$ (i.e., $\sigma_{{\rm H}\alpha+{\rm UV}}$) from the observed scatter ($\sigma_{\delta,{\rm obs}}$), the intrinsic scatter is less than $\sigma_{\delta} \lesssim 0.03$~dex for both mass bins.

Given the uncertainty in the error propagated to \deltaHa\ by the Balmer \avmass\ relation (Section~\ref{sec_avmass}), we also compute the intrinsic scatter in $\delta$ when employing only the duplicate sample.  We find that a smaller observed scatter in $\delta$ combined with a smaller uncertainty in \deltaHa\ (due to a negligible intrinsic scatter in the Balmer \avmass\ relation) leads to an intrinsic scatter in $\delta$ that is similar to that of the full sample ($\sim 0.03$~dex).  The uncertainty in the intrinsic scatter in the Balmer \avmass\ relation therefore does not affect this crucial measurement of $\sigma_{\delta}$.

Also shown in gray contours and histograms in the middle and right columns, respectively, are results from Monte Carlo simulations that assume all galaxies simply populate the SFS.  The simulated distributions shown include $0.3$~dex intrinsic scatter about the SFS added in quadrature to the bootstrapped uncertainties from the data in each column for \deltaUV\ and \deltaHa.  The overlap with the data reveals, again, how much of the observed scatter in  $\delta$ is explained by measurement uncertainties.

Our data limit the scatter between \deltaUV\ and \deltaHa\ for any proposed ensemble of IMG SFHs to $\lesssim 0.03$~dex ($1\sigma$). As we will discuss Section~\ref{sec_models}, this low number places strong limits on the time variability of SFHs---i.e., burstiness---for galaxies in this mass range.

\begin{figure*}
\epsscale{1.2}
\plotone{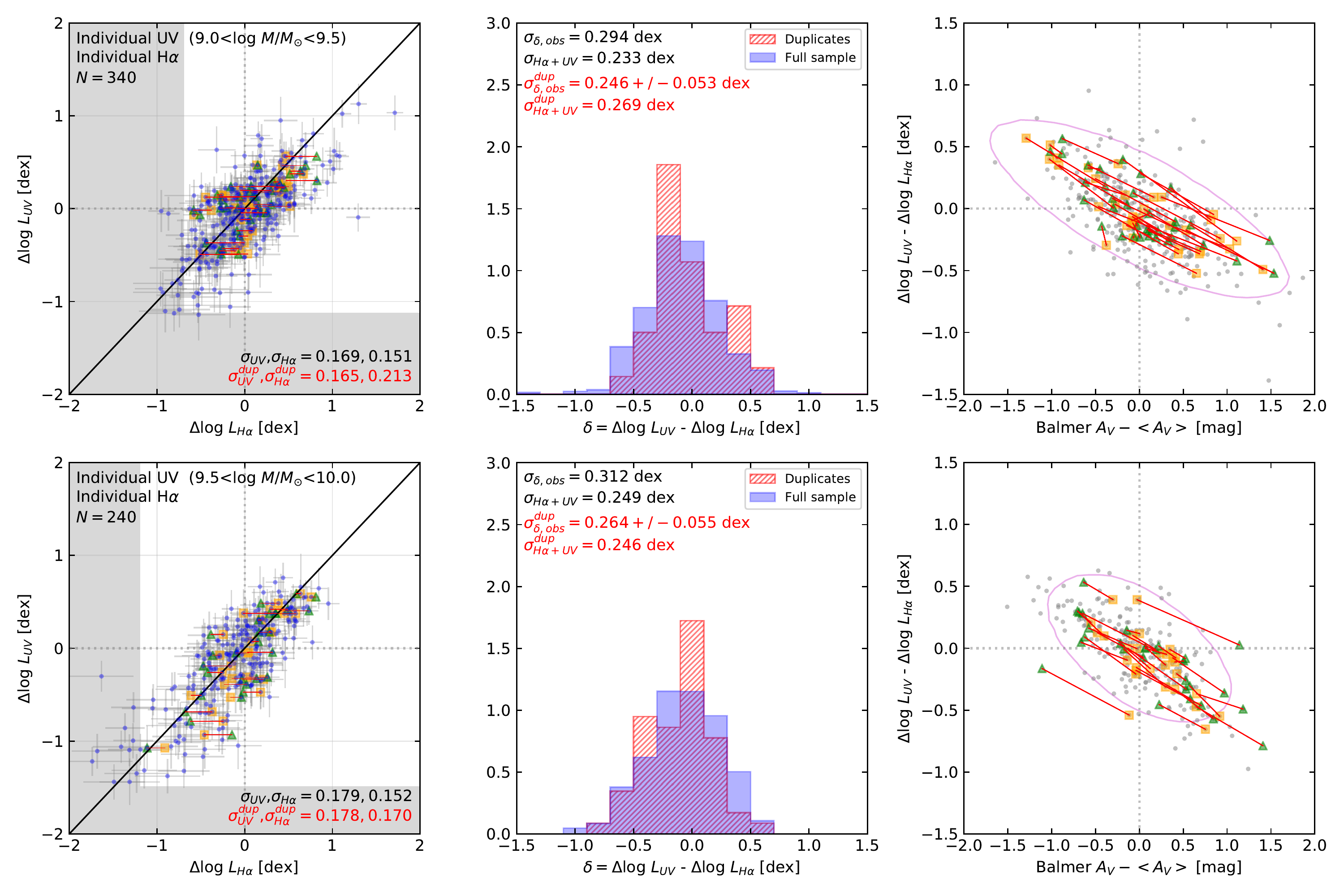}
\caption{({\em left}) Same as Figure~\ref{fig_HaUV_compare} but with \luv\ and \lalpha\ corrected for extinction using individually derived SED and Balmer $A_V$ values, respectively.  The sample of duplicates is indicated by green triangles and orange squares connected by red lines.  Median formal errors for the full sample are labeled.  Also labeled is the median \luv\ error from the duplicate sample ($\sigma_{UV}^{\rm dup}$) as well as the error in \lalpha\ inferred from the difference in duplicate measurements ($\sigma_{H\alpha}^{\rm dup}$).  The latter indicates that the formal error on \lalpha\ is significantly underestimated for both mass bins. ({\em middle}) Distribution of \deltauvha\ for the full sample and the duplicates.  While the typical error in $\delta$ for the full sample ($\sigma_{{\rm H}\alpha+{\rm UV}}$) is much smaller than the observed scatter, those errors are underestimated.  Meanwhile, the duplicate sample shows that the typical error in $\delta$ (i.e., $\sigma_{{\rm  H}\alpha+{\rm UV}}^{\rm dup}$) is within the uncertainty of the observed scatter. ({\em right}) $\delta$ vs. deviation of individually measured Balmer $A_V$ from the \avmass\ relation.  The duplicate measurements show that most of the observed scatter in $\delta$ is driven by measurement errors in Balmer $A_V$. Magenta contours show where 95\% of a simulated sample, constructed solely from measurement errors in Balmer $A_V$ and \luv, would lie.} \label{fig_HaUV_ind}
\end{figure*}

\subsubsection{\luv\ and \lalpha\ Corrected for Extinction with Individual SED and Balmer $A_V$ Values} \label{sec_delta_ind}

Figure~\ref{fig_HaUV_ind} shows \deltaUV\ vs. \deltaHa\ with each luminosity corrected for extinction using the individual SED and Balmer $A_V$ values, respectively.  The \deltaUV\ and \deltaHa\ values as well as detection limits are derived in a similar manner as for the sample in Figure~\ref{fig_delta_sfs}.  It should be noted that the sample sizes vary from Figure~\ref{fig_HaUV_compare} because slight variations in the fitted SFS impact the range of detectable \deltaUV\ and \deltaHa\ values (sample sizes given in the figures exclude objects below the detection limits).

The blue histograms in the middle column show that the observed scatter in $\delta$ is much larger when employing individual SED and Balmer $A_V$ corrections.  This is due to larger errors in the individual $A_V$ measurements that propagate into \luv\ and \lalpha.  The precise level of these luminosity errors is, however, fraught with uncertainty, especially for \lalpha.  When compared to our duplicate sample (pairs of squares and triangles connected with red lines), the formal errors given in the figure underestimate the range of random variations in \lalpha\ where $\sigma_{H\alpha}\sim 0.15$~dex for both mass bins.  We measure the error in \deltaHa\ for the duplicate sample by computing the scatter in the difference of pair \lalpha\ values (divided by $\sqrt{2}$) and find $\sigma_{H\alpha}^{dup}\sim 0.21/0.17$~dex for the low-/high-mass bins, which is much larger.

While an intrinsic scatter for $\delta$ can be computed for the full sample with the observed scatter and the formal errors, we know from the duplicate sample that the latter are underestimated.  Instead, employing only the duplicate sample, in which the error for \deltaHa\ is more reliably measured, we find that the total error in $\delta$, $\sigma_{H\alpha+UV}^{dup}$, is consistent with the observed scatter, $\sigma_{\delta,obs}^{dup}$, for both mass bins.  This would also be consistent with a small intrinsic scatter in $\delta$, as found in the previous section where extinction corrections were based on the \avmass\ trends.  However, we note that, given the smaller duplicate sample (35 and 29 pairs for the low- and high-mass bins, respectively), the bootstrapped errors for the observed scatter in $\delta$ are fairly large.

\subsubsection{The Case for Trend-corrected Luminosities as Our Fiducial Measurements} \label{sec_pref}

We have presented two different methods for correcting \luv\ and \lalpha\ for extinction: ({\em i}) using the individually computed SED and Balmer $A_V$ values, and ({\em ii}) using the latter to derive SED and Balmer \avmass\ trends that are applied to our sample.  Both methods produce results that are consistent with each other; however, there are several reasons to favor the trend-based results that we summarize here.

{\bf Smaller errors:} The trend-based results have smaller errors for \deltaUV\ and \deltaHa\ because the intrinsic scatters in the \avmass\ relations are smaller than the individual $A_V$ errors.  In fact, our sample of duplicates suggests that the formal errors for the individual Balmer $A_V$ measurements are underestimates of the true uncertainties, given additional sources of error (e.g., slit alignment, seeing variations, etc.).

{\bf Heteroscedasticity:} Uncertainties in $A_V$ from the trend-based method, especially for Balmer $A_V$, are more uniform across our sample, as the intrinsic scatter in the relation likely varies little within the IMG mass range \citep[see, e.g., ][]{garn2010b}.  In contrast, errors for individual $A_V$ measurements vary with a range of properties (e.g., $A_V$ value) and are also prone to outliers arising from data issues despite mitigation efforts (e.g., skylines contaminating \halpha\ flux).  Intrinsic scatters and errors added in quadrature are therefore more reliably interpreted with the trend-based method.

{\bf Residual correlations:}  Our data show residual trends for individual Balmer $A_V$ values based on their deviation from the \avmass\ relation.  In general, quantities related to aperture variations (e.g., redshift, total flux correction from the color aperture, galaxy size) contribute most to introducing additional observational scatter to \lalpha\ and therefore $\delta$.  These sources of uncertainty are unaccounted for in our individual Balmer $A_V$ measurements and again highlight the utility of the trend-based extinction corrections, which subsume these factors into the average trend.

\begin{figure*}
\epsscale{1.2}
\plotone{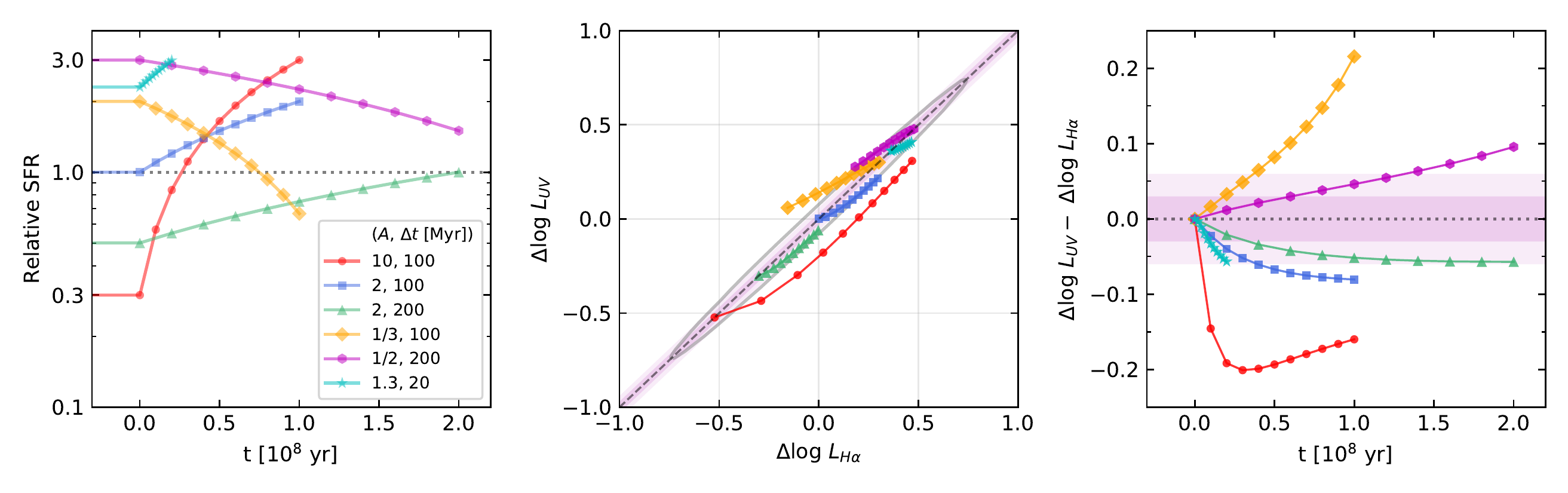}
\caption{Evolution of model SFHs in \deltaHa\ and \deltaUV. ({\em left}) SFHs relative to the SFS (dotted line).  SFHs linearly fall or rise after $t=0$~Myr with varying amplitudes ($A$) and durations ($\Delta t$). ({\em middle}) Corresponding evolution in \deltaHa\ and \deltaUV\ for $t\ge 0$~Myr (all models start on the one-to-one line). ({\em right}) Evolution in the ratio of the two luminosities.  For reference, the dark and light purple shaded regions represent our measured $1\sigma_{\delta}$ and $2\sigma_{\delta}$ scatter between the two luminosities, and the gray contour shows a basic model that represents where $95\%$ of IMGs would lie in the absence of errors.  The data rule out strong variations in recent SFH such as the strongly bursting (e.g., red circles) or quenching (orange diamonds) tracks.  A factor of $\sim 2$ change in SFR over $\sim 200$~Myr (e.g., green triangles, magenta hexagons) fits within the limits of the observations.  On shorter timescales, up to a $\sim 30\%$ change over $\sim 20$~Myr (e.g., cyan stars) is allowed by the data.}   \label{fig_schematic_SFH}
\end{figure*}

Finally, we examine whether the trend-based $A_V$ values bias the corrected luminosities when used in place of the individual $A_V$ values.  We focus on the Balmer $A_V$ values for this analysis.  The right column of Figure~\ref{fig_HaUV_ind} shows $\delta$ (with luminosities corrected from the individually measured SED and Balmer $A_V$ values) vs. the deviation in Balmer $A_V$ from its \avmass\ relation (i.e., \avdev).  Some galaxies have individual Balmer $A_V$ values that could be considered outliers from the \avmass\ relation, and these generally produce larger differences between \deltaUV\ and \deltaHa. However, as seen from the duplicate sample, most of the observed scatter in $\delta$ is dominated by measurement errors in Balmer $A_V$; values that are seemingly outliers from the \avmass\ relation in one measurement are not so in another.  In fact, among duplicates beyond $1$~mag of the \avmass\ relation, none have a pair measurement that lies even further from the relation.  Instead, the pairs are found reverting toward the relation as one might expect from the normally distributed residuals about the trend (Figure~\ref{fig_avmass}).  True outliers at such deviations are therefore likely to be rare.  In this context, the outlying individual $A_V$ measurements, scattered to large deviations from the \avmass\ relation due to errors, would produce corrected luminosities that are more discrepant from the truth when compared to corrections from the trend-based $A_V$ values.

To further emphasize the dominance of large uncertainties in $\delta$ from the individually derived $A_V$ corrections, we simulate a sample formed solely from measurement errors in Balmer $A_V$ (inferred from the duplicate sample in each mass bin; propagates into \lalpha\ and therefore $\delta$) as well as bootstrapped errors from \luv.  The magenta contours in Figure~\ref{fig_HaUV_ind} contain 95\% of the simulated samples and account for the bulk of the extent of the full sample in each mass bin.

Given the arguments above in favor of the trend-based $A_V$ corrections, we proceed with the corresponding measurements for $\delta$ but reiterate their consistency with the individual $A_V$ based results.

\section{Constraints on Fluctuating SFH\lowercase{s}: Insights from Models} \label{sec_models}

Fluctuating SFHs can produce a range of \deltaHa\ and \deltaUV\ values.  To better interpret the distribution of observed measurements, we study various model SFHs.  We use solar metallcity BC03 SPS models with \citet{chabrier2003} IMF to generate \luv\ and \lalpha\ for a given SFH.  In general, the effects of stellar metallicity on our results are negligible.  \luv\ is computed from the models using the same rest-frame 2800~\AA\ filter used for the observations.  \lalpha\ is computed from the ionizing photon rate (i.e., $Q(H^0)$) and assumes Case B recombination:
\begin{eqnarray}
\log L_{{\rm H}\alpha}=\log Q(H^0) -11.87
\end{eqnarray}

\begin{figure*}
\epsscale{1.2}
\plotone{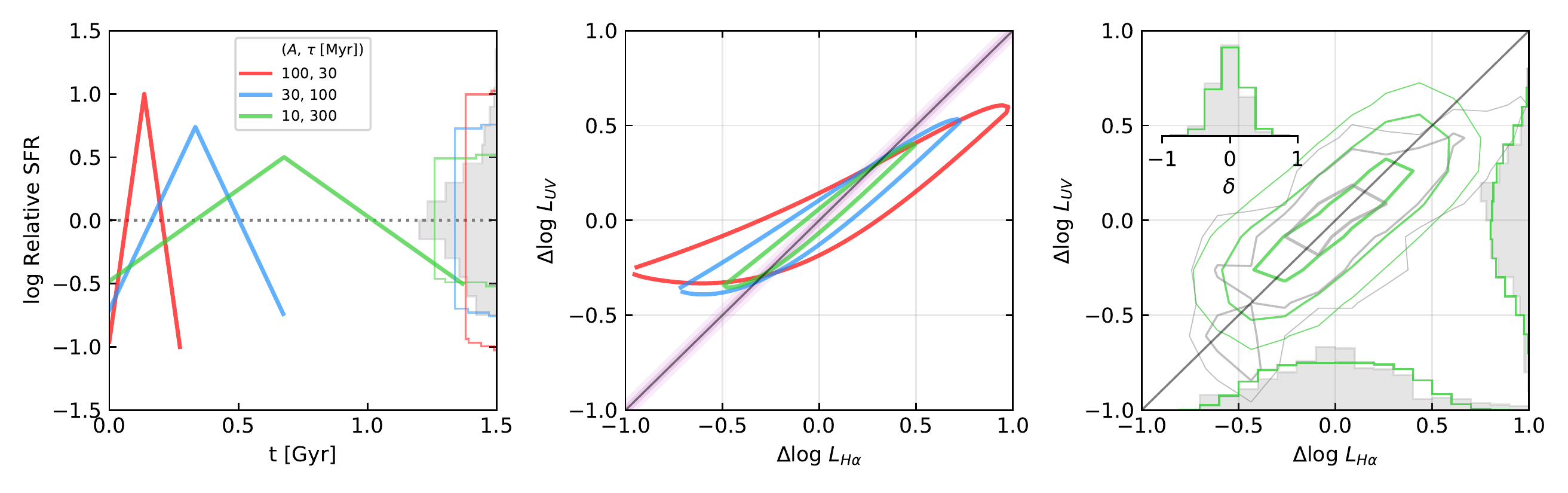}
\caption{({\em left}) Exponential burst SFHs with varying amplitudes ($A$) and $e$-folding times ($\tau$), normalized relative to the SFS (dotted line).  The models range from short bursts with large amplitudes (red) to more gradual bursts with small amplitudes (green).  Histograms show the resulting uniform SFR distribution (in log space) in comparison to the log-normal distribution of \deltaHa\ for SFGs in our lower-mass IMG sample (gray).  ({\em middle}) \deltaUV\ vs. \deltaHa\ for the model SFHs.  For reference, the dark and light purple shaded regions represent $1\sigma_{\delta}$ and $2\sigma_{\delta}$ ($\sigma_{\delta}=0.03$~dex).  Large-amplitude bursts (e.g., red and blue curves) are easily ruled out by our observations. ({\em right})  Two-dimensional distribution for the green model with measurement errors (bottom right) folded in.  Contours show the 25th, 75th, and 95th percentiles of the distribution.  The joint and individual distributions are too flat compared to the observations (gray contours).  No single exponential burst model appears to be characteristic of IMGs.}   \label{fig_singleSFH}
\end{figure*}

\subsection{Linearly Rising/Declining SFHs}

First, we examine a simple SFH that rises or falls linearly with time, to illustrate how different regions of \deltaUV\ vs. \deltaHa\ parameter space can be populated.  In Figure~\ref{fig_schematic_SFH}, the models begin their rise or fall at $t=0$ following a $2$~Gyr period of constant SFR.  Each model is defined by a duration ($\Delta t$) and amplitude ($A$), with the latter specified relative to the SFR at $t=0$.  All SFHs are normalized relative to a galaxy on the SFS (dotted line).  Each model spans $11$ equally spaced time steps at $t \ge 0$ with lengths of $\Delta t/10$ (note that $\Delta t$ may vary between different models).  The middle panel shows the evolution at $t \ge 0$ in \deltaUV\ and \deltaHa, which are computed from the models as follows:

\begin{eqnarray}
  \Delta\log L_X &=& \log\big(L_X(t)/L_X(t=0)\big) \\
  X &\in& \{{\rm UV},{\rm H}\alpha\} \nonumber
\end{eqnarray}
$L_X(t=0)$ represents the luminosity of the SFS in either UV or \halpha.  Its absolute value is arbitrary as \deltaUV\ and \deltaHa\ are relative quantities.  Because we are interested in changes in these two values over short timescales, the constant SFR assumed for the SFS is sufficient.  Note that all of the model tracks begin on the one-to-one relation, as they have a constant SFR prior to $t=0$.  The panel on the right shows the evolution in the ratio of the two luminosities (i.e., $\delta$).

The red SFH (circles) represents a galaxy starting off below the SFS that undergoes a factor of ten increase in SFR over $100$~Myr.  These fluctuations are of the scale seen in the FIRE simulation \citep[e.g.,][]{elbadry2016}.  As the SFR increases, the more sensitive \halpha\ tracer causes \deltaHa\ to increase more quickly than \deltaUV, leading the track to move off and to the right of the one-to-one line.  The blue SFH (squares), with the same duration but an amplitude five times smaller, deviates less significantly from the relation.  With the same amplitude as the latter but twice the duration, the green SFH (triangles) tracks even closer to the relation as the extended timescale dampens the impact of the changing SFR.  Meanwhile, both the orange and magenta SFHs (diamonds and hexagons, respectively) start out above the SFS and then decline by varying degrees.  As \halpha\ is quicker to respond, the model tracks move toward lower \deltaHa\ values and populate regions above the one-to-one line.  Finally, the cyan SFH (stars) shows a fluctuation over a much shorter duration ($20$~Myr).

For comparison, the shaded purple regions show our measured scatter between the two star formation indicators (i.e., $1\sigma_{\delta}$ and $2\sigma_{\delta}$).  In addition, a simple model representation of the observations is given by the gray contour in the middle panel, which shows where $95\%$ of IMGs would lie in the diagram in the absence of errors.  The model is produced by convolving the intrinsic scatter in the SFS with $\sigma_{\delta}=0.03$~dex scatter between \deltaUV\ and \deltaHa.

Our observations can rule out strong fluctuations in recent SFR, some of which are depicted in Figure~\ref{fig_schematic_SFH}.  Strong and frequent bursting (e.g., red circles) or quenching (e.g., orange diamonds) activity cannot be commonplace among IMGs, as their tracks deviate significantly from the range of allowable \deltaUV\ and \deltaHa\ values.  Even frequent factor of $\sim 2$ fluctuations over $100$~Myr timescales (blue squares) appear too strong when compared to the observations.

With these simple linear SFH models, we estimate that the data can accommodate fluctuations for the IMG population of at most a factor of $\sim 2$ over a duration of $200$~Myr and $\sim 30\%$ over shorter durations of $20$~Myr.

\begin{figure*}
\epsscale{1.2}
\plotone{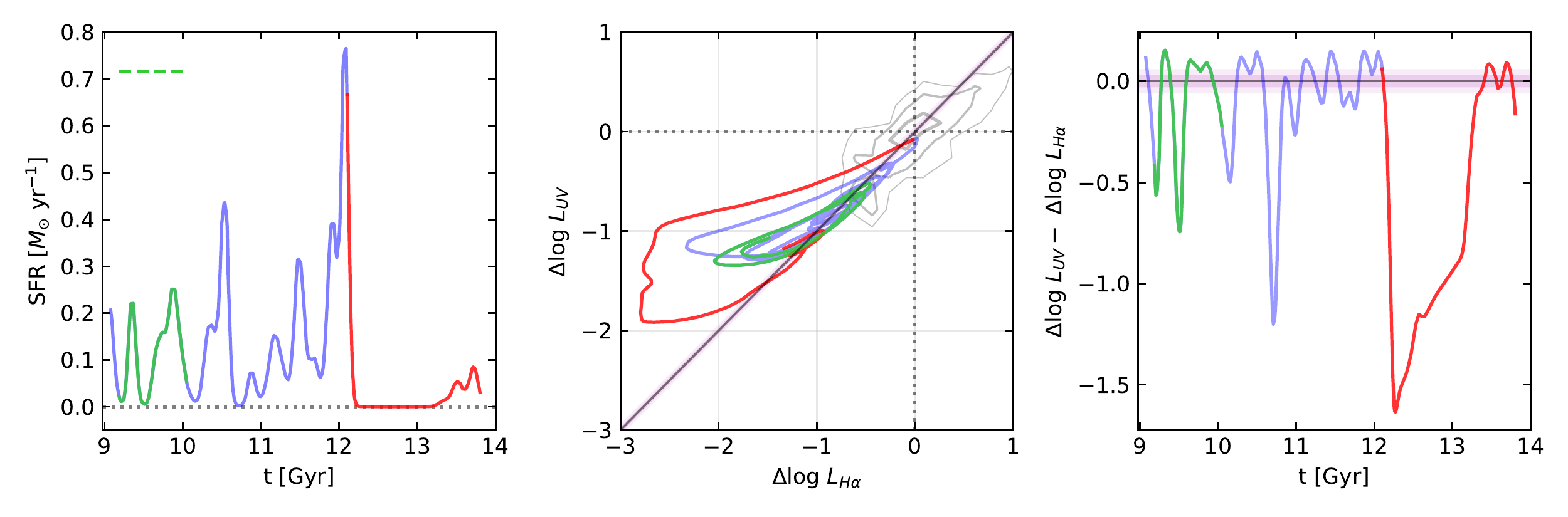}
\caption{({\em left}) SFH of the FIRE simulated IMG, m11, from \citet{elbadry2016} averaged over 100~Myr (the instantaneous SFR would vary more strongly than what is shown here).  Green portion indicates \zwindow, which would encompass our observations; however, the SFH behavior is qualitatively similar at later times.  The green dashed line indicates the median SFR for galaxies in our data with similar mass to m11.  ({\em middle}) Evolution in \deltaUV\ vs \deltaHa.  Overall low SFRs put m11 well below the observed SFS where most of the data lie (gray contours).  ({\em right}) Ratio of UV to \halpha\ luminosities (i.e., \deltauvha) as a function of time.  Strong variations in SFR lead to large departures from the one-to-one relation (i.e., \deltauvha~$=0$), which cannot be reconciled with our observations (shaded regions show $1\sigma_{\delta}$ and $2\sigma_{\delta}$ scatter between \deltaUV\ and \deltaHa).}   \label{fig_FIRE}
\end{figure*}

\subsection{A Characteristic SFH for IMGs is Unlikely}

Recent predictions of semi-periodic bursty SFHs \citep[e.g.,][]{elbadry2016,hopkins2018c} have motivated observational studies in identifying characteristic SFHs for galaxies in different stellar mass bins, including IMGs \citep{emami2019}.  We examine the exponential burst models from the latter work in the context of our observation.  These models ($\phi(t)=e^{t/\tau}$ over $0<t<D$ and $\phi(t)=e^{-(t-2D)/\tau}$ over $D<t<2D$) are defined by any two of the following three parameters: $e$-folding time ($\tau$), duration ($D$), and amplitude ($A$) (i.e., $D=\tau\log_{e}A$).  When normalized by their mean logarithmic SFR over $2D$ (or equivalently, their median SFR), they exhibit symmetry about the SFS in logarithmic space.  Figure~\ref{fig_singleSFH} shows example exponential burst models and their resulting tracks in \deltaHa\ and \deltaUV.  These are computed relative to the median value of \lalpha\ and \luv\ over a burst cycle, respectively, thereby making them comparable to our observations.  As the SFHs rise, the tracks in middle panel move along the bottom of the loop toward higher \deltaHa\ values.  As they peak and decline, the tracks move along the top toward lower \deltaHa\ values.  Larger amplitudes cause larger deviations from the one-to-one relation, while longer $\tau$ values cause the loops to rotate counterclockwise into closer alignment with the relation as gradual changes allow the UV and \halpha\ to equilibrate.  The SFHs are assumed to be periodic with only the first cycle plotted in the figure.

The IMG \halpha\ and UV measurements presented in \citet{emami2019} are consistent with those extracted from our entirely independent dataset.  While the observed scatter in the ratio of \lalpha\ to \luv\ for IMGs in their work is $\sim 0.08$~dex, the typical measurement error of $\sim 0.12$~dex suggests an intrinsic scatter near zero, similar to our findings.  With almost a factor of $\sim 30$ more IMGs, our data allow us to map the joint distribution of \lalpha\ and \luv\ with greater precision and test the hypothesis that an exponential burst model can serve as a characteristic SFH.

Our observations reveal challenges to the hypothesis of a characteristic SFH. (1) While the distribution of SFRs produced by any single model is symmetric in log space about the SFS, they are uniformly distributed as seen with the inset histograms in Figure~\ref{fig_singleSFH}({\em left}) (models have not been convolved with errors in this panel).  In contrast, the \deltaHa\ distribution (as close to an instantaneous SFR as possible) for our lower-mass sample of IMG SFGs (gray) shows the typically observed log-normal form \citep[see, e.g.,][]{popesso2019}.  (2)  While models with large amplitudes (e.g., red curve) can easily be ruled out for IMGs when comparing to the intrinsic scatter between \deltaUV\ and \deltaHa\ (middle panel), smaller-amplitude and longer-duration models (e.g., green curve) can produce values that are closer to the range allowed by our observational constraints. In the right panel, we investigate the joint distribution produced by the green model further.  Because the model is assumed to be characteristic of a population of galaxies, we sample it at equally spaced time intervals and inject errors that are typical of our observations in order to produce \deltaUV\ and \deltaHa\ values that we can compare to our observations.

The primary obstacle for the green model is that the SFH does not spend sufficient time close to the mode of the SFS, leading to a joint distribution that is not as peaked as the observations at \deltaHa~$=$~\deltaUV~$=0$.  Longer timescales ($\tau>300$~Myr) to fit the IMG distribution, as suggested in \citet{emami2019}, do not resolve the issue, as the individual \halpha\ and UV distributions only become flatter.  While smaller amplitudes may redistribute data into the peaks, they do so at the expense of the tails of the distribution.

While we have explored cyclical exponential burst models as a case study, it is likely that any single characteristic periodic or semi-periodic SFH will face similar issues.  Instead, a spectrum of fluctuation timescales is likely required in order to explain the data \citep[e.g., ][]{tacchella2020}.  Alternatively, an ensemble of stochastic SFHs may produce better agreement with observations \citep{kelson2014}.  Testing such nonparameterized SFH models will be addressed in a future paper.  In the next section, we instead focus on a recent hydrodynamical simulation that we can compare directly with our data.

\subsection{Comparison to an IMG in the FIRE Simulations}

So far, we have compared our observations to parameterized model SFHs.  In this section, we demonstrate how our data constrain SFH predictions from hydrodynamic simulations that incorporate feedback to regulate star formation.  Figure~\ref{fig_FIRE} shows the SFH of a simulated galaxy from FIRE \citep{hopkins2014} with a stellar mass at $z \sim 0$ of $10^{9.3}$~\msun\footnote{We inferred SFR from SSFR in \citet{elbadry2016} using a constant factor of $0.6$ for stars remaining due to mass loss from stellar evolution.  This assumption has no bearing on our conclusions.} \citep[named m11 in][]{elbadry2016}.  The SFH presented in that work was averaged over $100$~Myr, and thus the impact from rapid SFR perturbations are significantly diminished.  Nevertheless, even the variations predicted at these---and longer---timescales can be tested with our data.  In general, m11's SFH at late times undergoes semi-periodic fluctuations due to stellar feedback that drives gas and dark matter outward, only to cool and fall back into the center, reigniting the cycle \citep{elbadry2016}.  While m11 is just one galaxy, \citet{elbadry2016} suggest that it is representative of its mass range, such that different phases of its evolution can account for the observed scatter in the size-mass relation \citep[e.g.,][]{vanderwel2014}.

To generate the evolution in \deltaUV\ and \deltaHa\ the model SFH has been normalized by an SFR of $\sim 0.7$~\sfr\ (green dashed line).  This represents the median value of our sample with mass $10^{9} \leq M/M_{\odot} \leq 10^{9.3}$ (i.e., the mass range of m11 at \zwindow) and is also close to the value found by \citet{whitaker2012b}.  Note that this SFR is much higher than almost all of m11's late-time SFH.  As a result, the model tracks in \deltaUV\ and \deltaHa\ are displaced well below the origin (i.e., the peak of the SFS), where the bulk of the observations lie.   The galaxy was selected for zoom-in based on its isolation from other more massive halos, and thus environmental factors are unlikely to explain its overall low SFRs.  Shifting focus to the scatter about the one-to-one relation, the model is confronted with additional challenges when compared to our observations.  In particular, FIRE's feedback-induced quenching leads to departures above the one-to-one relation well beyond the measured intrinsic scatter between \deltaUV\ and \deltaHa.  In addition, the deviation of the bursting phase of the model from what the observations allow is likely understated, given that the FIRE SFH has been smoothed.  Ignoring the long quenching event (red) at late times ($t \gtrsim 9$~Gy), m11 spends $\sim 41\%$ of the time beyond the $\pm 2\sigma_{\delta}$ range between \deltaUV\ and \deltaHa.  This percentage is a lower limit, given the smoothing of the SFH.

We conclude that the cycle of burst and quench in FIRE---for IMGs in our mass range---is not supported by our observations, at least for SFH fluctuations of the strength and timescales presented in \citet{elbadry2016}.  This discrepancy, and deficiency of the FIRE simulation with regard to m11, may very well impact other findings related to strong feedback, such as the formation of cored dark matter density profiles \citep[e.g.,][]{chan2015,burger2022}.

\section{Discussion} \label{sec_discussion}

\subsection{Uncertainties}

The interpretation of our results depends on the \luv\ and \lalpha\ measurement uncertainties.  Given that we are operating in a regime where the scatter between the two luminosities is almost entirely dominated by observational errors, our interpretation of the data would change only if the error bars were overestimated.  We believe that our errors are, if anything, potentially underestimating more complex sources of scatter that can be envisioned and considered operable.  For example, differences in the apertures used to extract \luv\ and \lalpha\ could contribute additional scatter if those quantities exhibit strong spatial variations.  UV photons from older stellar populations contribute in some amount to \luv.  Also, the model SFHs discussed here have been derived assuming a single metallicity (solar), but variations in stellar population metallicity from galaxy to galaxy would affect the ratio of \lalpha\ to \luv\ \citep[see, e.g.,][]{lee2009b}.  Though the effects are likely minimal, alternative sources of ionizing radiation that produce \halpha\ flux, such as diffuse ionized gas \citep[][]{reynolds1973,oey2007,rsanders2017} or shock-heating \citep{mortazavi2019}, can also contribute additional scatter between \lalpha\ and \luv.

\subsection{SFR Volatility in the Nearby Universe}

Our sample is drawn from a general field survey spanning a range of environments \citep[see, e.g.,][]{patel2018}, and barring cosmic variance, it should be representative of IMG populations at late times.  In the nearby Universe, many IMGs have been documented to undergo bursts of star formation that would exceed the limits of SFR variability determined in this work (factor of $\sim 2$ over $200$~Myr).  For example, \citet{chandar2021} recently measured more than a factor of 100 increase in the SFR of a post-starburst galaxy in under $100$~Myr.  However, this particular galaxy was notable for interacting with a neighbor.  This is often the case for other well-documented starburst galaxies, as environmental factors play a leading role in regulating their recent SFH \citep[e.g., M82;][]{mayya2006}.  Such objects may exist in our survey: for example, outliers in the Balmer \avmass\ diagram could have their \lalpha\ suppressed to lower values when adopting the median relation for dust corrections.  However, the outlier population by definition would be small, just like the observed population of post-starburst galaxies \citep[e.g.][]{wild2016}.  For the vast majority of IMGs, our analysis shows more limited SFR volatility.

\section{Summary} \label{sec_summary}

We studied fluctuations in the recent SFHs of IMGs (\massrange).  We compared \halpha\ and UV luminosities for a sample of \nsample\ IMGs in COSMOS at \zwindow.  The two luminosities probe SFRs on different timescales and together can reveal the recent SFH trajectory.  Magellan IMACS spectroscopy was obtained to measure \lalpha\ while rest-frame 2800~\AA\ photometry was used to compute \luv.

Our main conclusions are the following:
\begin{enumerate}
\item In determining dust corrections---by far the dominant source of error in this type of study---our duplicate spectroscopic observations reveal that the individually measured Balmer $A_V$ (used to correct \lalpha) have fairly high uncertainties ($\sigma_{A_V} \sim 0.40$-$0.54$~mag) the sources of which appears to be variations in slit alignment and other observational effects.  Instead, we find that the intrinsic scatter about the Balmer \avmass\ relation propagates a smaller uncertainty for Balmer $A_V$ ($\sigma_{A_V} \lesssim 0.36$~mag).  For SED $A_V$ (used to correct \luv), the propagated error from the SED~\avmass\ relation ($\sigma_{A_V}=0.20$~mag) also represents an improvement over the individually measured $A_V$ from SED fits ($\sigma_{A_V}=0.23$~mag).  We therefore use our derived \avmass\ relations to apply extinction corrections.
  
\item We compute residuals from the \halpha\ and UV star forming sequences (\deltaHa\ and \deltaUV).  Differences between the two quantities reflect variations in the recent SFHs.  Measurement uncertainties make it difficult to classify most galaxies with a rising or falling SFH.  We therefore focus on the scatter in the two luminosities for the ensemble population to constrain SFR fluctuations for the IMG population.  We find that measurement uncertainties explain most of the observed scatter between the two luminosities, leaving an intrinsic scatter of only $\sigma_{\delta} \lesssim 0.03$~dex.  At this level, a linearly rising or falling SFH spanning $200$~Myr would be limited to only a factor of $\sim 2$ change in SFR.  For shorter durations of $20$~Myr, only a $\sim 30\%$ change in SFR is allowed.

\item Any single SFH model, such as an exponential burst with a fixed timescale, is unlikely to characterize all IMG SFHs.  Instead, a diversity of SFHs is likely needed to describe the data.  Among the spectrum of fluctuation timescales permitted, the power would be substantially diminished at short (emission-line) timescales as compared to the longer timescales probed by starlight.

\item Predictions of strong SFR fluctuations on $\lesssim 100$~Myr timescales seen in hydrodynamic simulations \citep[e.g.,][]{elbadry2016} of IMGs are inconsistent with our observations.  Stellar feedback is coupled to the SFHs in these simulations and is responsible for driving numerous galaxy-scale properties, including morphological structure and dark matter density profiles.  The strength of such feedback should be re-examined, given our tight constraints on recent SFR fluctuations.

\end{enumerate}

% ############################################################

\begin{acknowledgments}
\indent We thank the anonymous referee for providing insightful comments that helped improve this work. S.G.P. acknowledges Sage and Cypress Patel.
\end{acknowledgments}

%\bibliography{all}
%\bibliographystyle{apj}

\end{document}